\documentstyle[aps,preprint,amssymb,epsfig]{revtex}
\input prepictex
\input pictex
\input postpictex

\begin{document}


\def\Bq{B_q^0}
\def\Bd{B_d^0}
\def\Bs{B_s^0}
\def\Bds{B_{d,s}^0}
\def\Bbar{{\bar B}}
\def\Bbarq{{\bar B}_q^0}
\def\Bbard{{\bar B}_d^0}
\def\Bbars{{\bar B}_s^0}
\def\Bbards{{\bar B}_{d,s}^0}

\def\rhobar{{\bar\rho}}
\def\etabar{{\bar\eta}}
\def\VMA{\gamma^\mu(1-\gamma_5)}
\def\VmA{\gamma_\mu(1-\gamma_5)}

\def\epjc#1#2#3{Eur. Phys. J. C {\bf #1}, #3 (#2)}
\def\ijmpa#1#2#3{Int. J. Mod. Phys. A {\bf #1}, #3 (#2)}
\def\mpl#1#2#3{Mod. Phys. Lett. A {\bf #1}, #3 (#2)}
\def\npb#1#2#3{Nucl. Phys. {\bf B#1}, #3 (#2)}
\def\plb#1#2#3{Phys. Lett. B {\bf #1}, #3 (#2)}
\def\prd#1#2#3{Phys. Rev. D {\bf #1}, #3 (#2)}
\def\prl#1#2#3{Phys. Rev. Lett. {\bf #1}, #3 (#2)}
\def\rep#1#2#3{Phys. Rep. {\bf #1}, #3 (#2)}
\def\zpc#1#2#3{Z. Phys. {\bf #1}, #3 (#2)}


\title{Non-factorizable effects on $\Bbar^0\to D^{(*)0}\pi^0$}
\author{Jong-Phil Lee\footnote{e-mail: jplee@phya.yonsei.ac.kr}}
\address{Department of Physics and IPAP, Yonsei University, Seoul, 120-749, Korea}

\tighten
\maketitle

\begin{abstract}
Non-factorizable effects on the color-suppressed $B\to D^{(*)}\pi$ decay modes
are analyzed.
Recent observations of $\Bbar^0\to D^{(*)0}\pi^0$ by Belle and CLEO 
strongly suggest that there exists a non-zero strong phase difference between 
color-allowed and color-suppressed decay modes, and the 
factorization parameter $a_2$ associated with the color-suppressed decay mode
is process dependent.
In the heavy quark limit where $b$ and $c$ are heavy, the process dependence of 
$a_2(D\pi,D^*\pi)$ is due to the different configuration of the heavy quark
spin relative to the light degrees of freedom.
From the experimental data, 
the heavy quark spin symmetry breaking contributions to the non-fatorizable
effects are estimated to be $23-28\%$.
\end{abstract}
\pagebreak


With the beginnig of the $B$-factory era, a lot of exciting data are waiting 
for reliable theoretical explanations.
Nonleptonic two-body decays, among them, possess abundunt phenomena including
the famous $B\to J/\psi K$.
In a theoretical point of view,
the tow-body hadronic decays are quite difficult to deal with
because of our poor understandings of the nonperturbative effects on the 
hadronic matrix elements.
The most widely used method is the factorization assumption in which the 
hadronic matrix element of the four-quark operator is described by the product 
of two current matrix elements.
\par
There are important parameters $a_i$ engaged in the factorization.
If the factorization were exact, then $a_i$ were just the linear combinations
of the Wilson coefficients of the effective Hamiltonian.
The index $i$ of $a_i$ is related to the classification category of the 
nonleptonic two-body decays.
We follow the usual convention, where $a_1$ is responsible
for color-allowed external $W$-emission while $a_2$ for color-suppressed 
internal $W$-emission amplitudes.
\par
Recent progress in theory of nonleptonic $B$ decays includes the QCD
improvements \cite{Beneke,Neubert2,Chay}.
By incorporating the hard-scattering effects, it is possible to calculate the
non-factorizable radiative corrections.
The values of $a_1$ for $\Bbar^0\to D^{(*)+}L^-$ where $L$ is a light meson,
calculated in this way, show the near universailty, accomodating the 
experimental data.
The process-dependent contributions turn out to be small.
On the other hand, there exist difficulties in calculating $a_2$.
To apply the same method one should assume that the charm quark is light which
is not a good approximation.
\par
Experimentally, Belle and CLEO reported the first observation of 
$\Bbar^0\to D^{(*)0}\pi^0$ \cite{Belle,CLEO}.
This process corresponds to the class-II (''color-suppressed'') where the 
final states are neutral mesons, and the decay amplitude is proportional to 
$a_2$.
Because the nonfactorizable effects appear mainly in $a_2$ rather
than $a_1$, the new data will check the validity of the factorization
hypothesis. 
The implications of the data in this direction are discussed in recent papers.
The new experimental data strongly suggest that there exists a non-zero 
strong-phase difference between $a_1$ and $a_2$ \cite{Xing,Cheng,Neubert1}.
In addition, new measurements result in the first verification of the process 
dependence of $a_2$.
Typical value of $a_2$ from other processes yields a very small branching ratio
compared to the recent data.
One dilema involved is that $a_2$ cannot be too large to fit the new data
because a large value of it will increase the branching fraction of the 
class-III decay mode $B^-\to D^{(*)0}\pi^-$, producing another discrepancy.
The observed relative strong phases work well to satisfy both requirements.
In short, new experimental data disfavor the (naive) factorization hypothesis.
Non-factorizable effects will play crucial roles in the color-suppressed decay
modes.
\par
In this paper, we review the implications of recent experimental data on
$\Bbar^0\to D^{(*)0}\pi^0$, and extract the process-dependent 
non-factorizable effect $\epsilon_{\rm NF}$ on $a_2$ from the data.
A special attention is paid to the ratio 
$\epsilon_{\rm NF}^{D^{*0}\pi^0}/\epsilon_{\rm NF}^{D^0\pi^0}$.
In the heavy quark limit, the final states are distinguished by the heavy
quark spin configuration relative to the light degrees of freedom.
Since the heavy quark spin symmetry is broken by the subleading chromomagnetic
interactions, the ratio will measure this kind of corrections in the heavy 
quark mass expansion.
\par
Let us fist summarize the implications of the recent measurements by Belle and
CLEO.
The effective Hamiltonian for $b\to c\bar{u}d$ is
\begin{equation}
{\cal H}_{\rm eff}=\frac{G_F}{\sqrt{2}}V_{cb}V^*_{ud}\Big[
 c_1(\mu)(\bar{d}u)_{V-A}(\bar{c}b)_{V-A}
 +c_2(\mu)(\bar{c}u)_{V-A}(\bar{b}d)_{V-A}\Big]~,
\end{equation}
where $(\bar{q_i}q_j)_{V-A}=\bar{q_i}\VMA q_j$, and $c_i$ are the Wilson
coefficients.
After the proper Fierz transformations, the decay amplitudes of $\Bbar\to D\pi$
are given by
\begin{mathletters}
\label{amp}
\begin{eqnarray}
{\cal A}_{+-}&\equiv&
  {\cal A}(\Bbar^0\to D^+\pi^-)={\cal T}+{\cal E}~,\\
{\cal A}_{00}&\equiv&
  {\cal A}(\Bbar^0\to D^0\pi^0)=\frac{1}{\sqrt{2}}(-{\cal C}+{\cal E})~,\\
{\cal A}_{0-}&\equiv&
  {\cal A}(B^-\to D^0\pi^-)={\cal T}+{\cal C}~,
\end{eqnarray}
\end{mathletters}
where 
\begin{mathletters}\label{topology}
\begin{eqnarray}
{\cal T}&=&\frac{G_F}{\sqrt{2}}V_{cb}V^*_{ud}
 \langle\pi^-|(\bar{d}u)_{V-A}|0\rangle
 \langle D^+|(\bar{c}b)_{V-A}|\Bbar^0\rangle a_1 \nonumber\\
&=&i\frac{G_F}{\sqrt{2}}V_{cb}V^*_{ud}(m_B^2-m_D^2)f_\pi F_0^{BD}(m_\pi^2)
 a_1~,\\
{\cal C}&=&\frac{G_F}{\sqrt{2}}V_{cb}V^*_{ud}
 \langle D^0|(\bar{c}u)_{V-A}|0\rangle
 \langle\pi^0|(\bar{d}b)_{V-A}|\Bbar^0\rangle a_2\nonumber\\
&=&i\frac{G_F}{\sqrt{2}}V_{cb}V^*_{ud}(m_B^2-m_\pi^2)f_D F_0^{B\pi}(m_D^2)
 a_2~,\\
{\cal E}&=&\frac{G_F}{\sqrt{2}}V_{cb}V^*_{ud}
 \langle\Bbar^0|(\bar{d}b)_{V-A}|0\rangle
 \langle D^0\pi^0|(\bar{c}u)_{V-A}|0\rangle a_2\nonumber\\
&=&i\frac{G_F}{\sqrt{2}}V_{cb}V^*_{ud}(m_D^2-m_\pi^2)f_B F_0^{0\to D\pi}(m_B^2)
 a_2~,
\end{eqnarray}
\end{mathletters}
are the color-allowed external $W$-emission, color-suppressed internal $W$-
emission and $W$-exchange amplitudes, respectively.
Note that (\ref{amp}) satisfies the isospin triangle relation
\begin{equation}
{\cal A}_{+-}=\sqrt{2}{\cal A}_{00}+{\cal A}_{0-}~.
\label{triangle}
\end{equation}
The weak form factor $F_0$ is defined by
\begin{equation}
\langle P_2(p')|V_\mu|P_1(p)\rangle=\Bigg[
 (p+p')_\mu-\frac{m_1^2-m_2^2}{q^2}q_\mu\Bigg]F_1(q^2)
 +\frac{m_1^2-m_2^2}{q^2}q_\mu F_0(q^2)~,
\end{equation}
with $q_\mu=p_\mu-p'_\mu$.
There are various apporaches to get the $q^2$ dependence of the form factors.
We adopt the Neubert-Rieckert-Stech-Xu (NRSX) model \cite{NRSX}, the 
relativisitic light-front (LF) quark model \cite{LF}, the Neubert-Stech model
\cite{NS}, and the Melikhov-Stech (MS) model \cite{MS}.
The decay constants are given by as usual
\begin{equation}
\langle P(p)|(\bar{q_i}q)_A|0\rangle=if_P p_\mu~.
\end{equation}
\par
\begin{figure}
\begin{center}
\epsfig{file=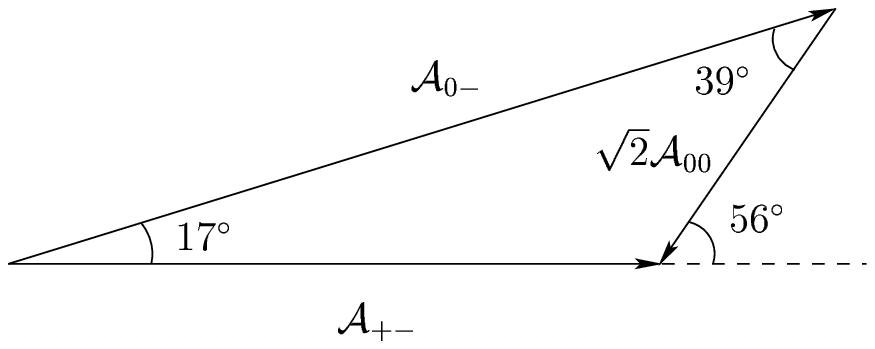}
\end{center}
\caption{}
\label{Dpi}
\end{figure}

From the experimental data \cite{Belle,CLEO,PDG},
\begin{eqnarray}
{\cal B}(\Bbar^0\to D^+\pi^-)&=&(3.0\pm0.4)\times10^{-3}~,\nonumber\\
{\cal B}(B^-\to D^0\pi^-)&=&(5.3\pm0.5)\times10^{-3}~,\nonumber\\
{\cal B}(\Bbar^0\to D^0\pi^0)&=&(0.27\pm0.05)\times10^{-4}~,\nonumber\\
\kappa\equiv\frac{\tau_{B^-}}{\tau_{\Bbar^0}}&=&1.073\pm 0.027,
\label{dataDpi}
\end{eqnarray}
one gets (with only central values)
\begin{equation}
\frac{\sqrt{2}{\cal A}_{00}}{{\cal A}_{+-}}=-0.42 e^{i56^\circ}~,~~~
\frac{\sqrt{2}{\cal A}_{00}}{{\cal A}_{0-}}=-0.33 e^{i39^\circ}~.
\end{equation}
In (\ref{dataDpi}), the value of ${\cal B}(\Bbar^0\to D^0\pi^0)$ is a combined
result of Belle and CLEO measurements.
The situation is depicted in Fig.\ \ref{Dpi}.
The ratio $\sqrt{2}{\cal A}_{00}/{\cal A}_{+-}$ is proportional to $a_2/a_1$ as
\begin{equation}
\frac{a_2}{a_1}=\Bigg(\frac{-\sqrt{2}{\cal A}_{00}}{{\cal A}_{+-}}\Bigg)
 \Bigg(\frac{m_B^2-m_D^2}{m_B^2-m_\pi^2}\Bigg)\Bigg(\frac{f_\pi}{f_D}\Bigg)
 \Bigg(\frac{F_0^{BD}(m_\pi^2)}{F_0^{B\pi}(m_D^2)}\Bigg)~,
\label{a2a1}
\end{equation}
where we have neglected the internal $W$-exchange diagram $\cal E$.
Using the NRSX model for the form factors, we have \cite{Cheng}
\begin{equation}
\frac{a_2}{a_1}=0.45 e^{i56^\circ}~.
\end{equation}
The results for other models are given in Table \ref{result}.
\par
It is quite convenient to introduce the isospin amplitudes.
The decomposition of the decay amplitudes into the isospin ones is given by
\begin{mathletters}
\begin{eqnarray}
{\cal A}_{+-}&=&\sqrt{\frac{2}{3}}A_{1/2}+\sqrt{\frac{1}{3}}A_{3/2}~,\\
{\cal A}_{00}&=&\sqrt{\frac{1}{3}}A_{1/2}-\sqrt{\frac{2}{3}}A_{3/2}~,\\
{\cal A}_{0-}&=&\sqrt{3}A_{3/2}~,
\end{eqnarray}
\end{mathletters}
where the coefficients are the Clebsch-Gordan, and the last expression comes
from the triangle relation (\ref{triangle}).
It is not difficult to see that
\begin{eqnarray}
|A_{1/2}|&=&|{\cal A}_{+-}|^2+|{\cal A}_{00}|^2-\frac{1}{3}|{\cal A}_{0-}|~,
\nonumber\\
|A_{3/2}|^2&=&\frac{1}{3}|{\cal A}_{0-}|~,\nonumber\\
\cos\delta&=&\frac{3|{\cal A}_{+-}|^2-2|A_{1/2}|^2-|A_{3/2}|^2}
 {2\sqrt{2}|A_{1/2}||A_{3/2}|}~,
\end{eqnarray}
where $\delta$ is the relative phase between $A_{1/2}$ and $A_{3/2}$.
From the experimental data for the branching ratios,
\begin{equation}
\frac{A_{1/2}}{A_{3/2}}=0.99 e^{i27^\circ}~.
\end{equation}
\par
\begin{figure}
\begin{center}
\epsfig{file=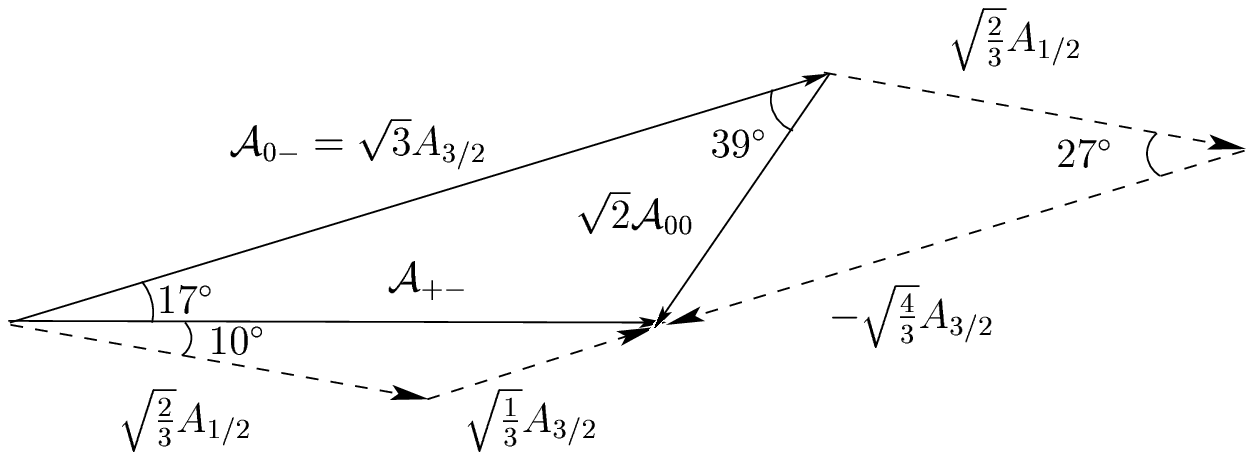}
\end{center}
\caption{}
\label{iso}
\end{figure}

All of the above results are encapsulated in Figs.\ \ref{Dpi},\ref{iso}.
A rather large value of ${\cal A}_{00}$ (or $a_2$) accomodates well with
the known value of ${\cal A}_{0-}$ via the relative strong phase 
$\approx 56^\circ$.
Figure \ref{iso} shows the isospin decomposition and the relative phase between
the isospin amplitudes.
It is clear that $|a_2/a_1|$ is greater than $a_2^{\rm eff}/a_1^{\rm eff}$.
Note that 
$|a_1|\sim |{\cal A}_{+-}|<|\sqrt{2/3}A_{1/2}|+|\sqrt{1/3}A_{3/2}|
\sim a_1^{\rm eff}$,
and 
$|a_2|\sim |\sqrt{2}{\cal A}_{00}|>|\sqrt{4/3}A_{3/2}|-|\sqrt{2/3}A_{1/2}|
\sim a_2^{\rm eff}$.
It is also expected that $||a_1|-a_1^{\rm eff}|<||a_2|-a_2^{\rm eff}|$.
Numerical results in \cite{Cheng} support this tendency, meaning that 
$a_2$ is more sensitive to the final-state interactions.
\par
\begin{figure}
\begin{center}
\epsfig{file=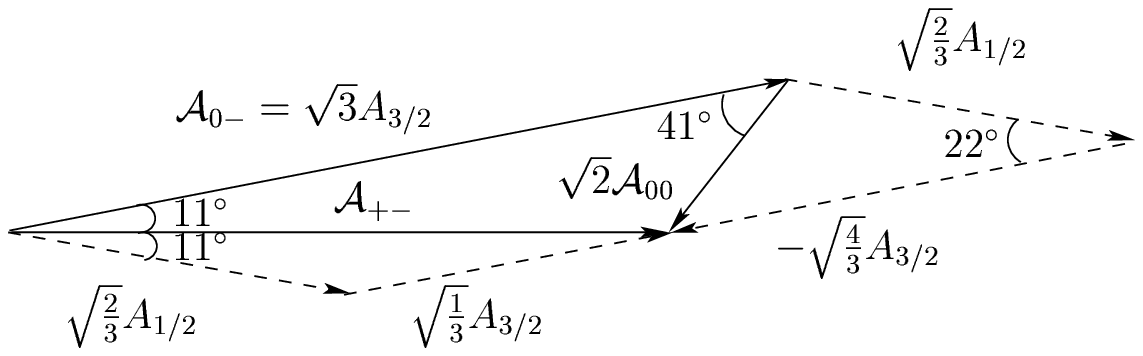}
\end{center}
\caption{}
\label{Dpi*}
\end{figure}

We can do the same analysis for $B\to D^*\pi$.
The branching ratios are
\begin{eqnarray}
{\cal B}(\Bbar^0\to D^{*+}\pi^-)&=&(2.76\pm 0.21)\times 10^{-3}~,\nonumber\\
{\cal B}(B^-\to D^{*0}\pi^-)&=&(4.6\pm 0.4)\times 10^{-3}~,\nonumber\\
{\cal B}(\Bbar^0\to D^{*0}\pi^0)&=&(1.7\pm 0.5)\times 10^{-4}~,
\end{eqnarray}
where the last one is a combined value of Belle and CLEO.
Using the ''tilde'' for the observables of $D^*\pi$, we have
\begin{mathletters}
\begin{eqnarray}
\frac{\sqrt{2}\tilde{\cal A}_{00}}{\tilde{\cal A}_{+-}}
&=&-0.35e^{i52^\circ}~,~~~
\frac{\sqrt{2}\tilde{\cal A}_{00}}{\tilde{\cal A}_{0-}}=-0.28e^{i41^\circ}~,\\
\frac{\tilde{a}_2}{\tilde{a}_1}&=&0.28 e^{i52^\circ}~~~{\rm for~NRSX}~,\\
\frac{\tilde{A}_{1/2}}{\tilde{A}_{3/2}}&=&1.02 e^{i22^\circ}~.
\end{eqnarray}
\end{mathletters}
Other values of $\tilde{a}_2$ corresponding to LF, MS, and NS are given in
Talbe \ref{result}.
The isospin triangle and its decomposition into the isospin amplitudes are
shown in Fig.\ \ref{Dpi*}.
\par
If the factorization assumption were exactly correct, then the factorization
parameters $a_i$ are real and there would be no phases between them.
In addition, $a_i$ were expected to be universal, i.e., process independent.
This is not the case of real world, as new experimental data strongly assert,
and non-factorizable effects play a significant role in $B\to D\pi$.
From the values of $a_2(D^{(*)}\pi)$, we can estimate the non-factorizable
effects.
Non-factorizable effects on $a_1$ turn out to be small \cite{NRSX,NS,Deandrea},
so we concentrate on $a_2$.
In general, the non-factorizable effects $\epsilon_{\rm NF}$ can be included in
$a_2$ as 
\cite{Cheng2}
\begin{equation}
a_2(D^{(*)}\pi)=c_2(\mu)
+\Bigg(\frac{1}{N_c}
+\frac{\epsilon_{\rm NF}^{B\pi,D^{(*)}}(\mu)}{f_{D^{(*)}}}\Bigg)c_1(\mu)~.
\end{equation}
The $\mu$-dependence of $\epsilon_{\rm NF}$ compensates that of $c_i(\mu)$ to
make $a_2$ $\mu$-independent.
We fix $\mu=m_b$.
The Wilson coefficients $c_i(m_b)$ can be obtained easily by the RGE 
\cite{NS}:
\begin{equation}
c_1(m_b)=1.132~,~~~~~ c_2(m_b)=-0.286~.
\end{equation}
\par
Note that $\epsilon_{\rm NF}$ is process dependent.
The process dependence of $a_2$ is attributed to that of $\epsilon_{\rm NF}$.
In a theoretical point of view, the ''process dependence'' is discouraging
news since it diminishes the predictive power.
As for $D^0\pi^0$ and $D^{*0}\pi^0$ in the final states, however, we can
relate $D^0$ and $D^{*0}$ using the heavy quark symmetry \cite{HQET}, assuming
that $m_c$ is heavy enough.
The ratio $|\epsilon_{\rm NF}^{B\pi,D^*}/\epsilon_{\rm NF}^{B\pi,D}|$ thus
can be understood in the context of the heavy quark effective theory.
Using the NRSX values for $F_0$, we have
\begin{equation}
R_{\rm NF}\equiv
\Bigg|\frac{\epsilon_{\rm NF}^{B\pi,D^*}}{\epsilon_{\rm NF}^{B\pi,D}}\Bigg|
=0.72~,
\end{equation}
where the decay constants $f_D\approx 200$ MeV and $f_{D^*}\approx 230$ MeV
are used.
Results from other models for the weak form factors are summarize in 
Table \ref{result}.
\par
In the heavy quark limit where $m_{b,c}\to \infty$, the light degrees of freedom
do not care about the heavy quark's spin configurations.
The heavy quark spin symmetry breaking occurs at ${\cal O}(1/m_Q)$, where 
$m_Q$ is the heavy quark mass.
The symmetry breaking is realized by the chromomagnetic interaction terms
in the effective Lagrangian at NLO.
Thus the value $R_{\rm NF}-1$ measures the heavy quark spin symmetry breaking effects.
It suggests that the symmetry breaking corrections give negative contributions,
and the enhancement is $\approx 23-28\%$ in magnitude.
\par
For one step further, we should implement the QCD improvement for $a_2$ or trace
out the sources of the non-factorizable effects.
Regarding the QCD improvments of $a_2(D\pi)$, however, there is no known 
systematics yet.
As pointed out in \cite{Beneke,Neubert2}, the QCD factorization formulae cannot
be directly applied to $a_2(D\pi)$ because the color-transparency arguments
break down when the emitted meson is heavy.
The discrimination of various sources of the non-factorizable effects is also
far from satisfaction.
The final-state interaction is a good candidate but the problem of inelasticity
in the rescattering remains unsolved yet \cite{Beneke,Neubert1}.
\par
In summary, we extract the non-factorizable effects on $a_2$ from the new
experimental data.
A large dependence of $a_2$ on the process certainly reduces the predictive
power.
In the context of the heavy quark symmetry, the heavy quark spin symmetry 
breaking contributions to $\epsilon_{\rm NF}^{B\pi,D^{(*)}}$ are estimated.
It still remains as a challenging work to disentangle various sources of the 
non-factorizable effects on $a_2$.

\vskip 1cm
\begin{center}
{\large\bf Acknowledgements}
\end{center}
JPL gives thanks to Sechul Oh for helpful discussions.
This work was supported by the BK21 Program of the Korea Ministry of Education.


\newpage

\begin{center}{\large\bf FIGURE CAPTIONS}\end{center}
\noindent
Figure 1
\\
Isospin triangle for $B\to D\pi$.
\vskip .3cm
\par

\noindent
Figure 2
\\
Decomposition into the isospin amplitudes for $B\to D\pi$.
\vskip .3cm
\par

\noindent
Figure 3
\\
Isospin triangle and its decomposition into the isospin amplitudes for 
$B\to D^*\pi$.
\vskip .3cm
\par

\vskip 3cm
\begin{center}{\large\bf TABLE CAPTIONS}\end{center}
\noindent
Table 1
\\
Numerical results for $a_2$ and the non-factorizable effects.
\vskip .3cm
\par

\pagebreak

\begin{table}
\caption{}
\begin{tabular}{c|cccc}
& NRSX & LF & MS & NS \\\hline
$F_0^{B\pi}(m_D^2)$ & 0.37 & 0.34 & 0.32 & 0.27 \\
$F_0^{BD}(m_\pi^2)$ & 0.69 & 0.70 & 0.67 & 0.63 \\
$a_2(D\pi)$ & 0.39$e^{i56^\circ}$ & 0.43$e^{i56^\circ}$ & 0.45$e^{i56^\circ}$ & 0.54$e^{i56^\circ}$ \\
$\tilde{a}_2(D^*\pi)$ & 0.26$e^{i52^\circ}$ & 0.30$e^{i52^\circ}$ & 0.32$e^{i52^\circ}$ & 0.36$e^{i52^\circ}$ \\
$|\epsilon_{\rm NF}^{B\pi,D}/f_D|$ & 0.31 & 0.34 & 0.36 & 0.44 \\
$|\epsilon_{\rm NF}^{B\pi,D^*}/f_{D^*}|$ & 0.19 & 0.22 & 0.24 & 0.28 \\
$R_{\rm NF}$ & 0.72 & 0.76 & 0.77 & 0.73 \\
\end{tabular}
\label{result}
\end{table}
\end{document}